# Needle beams and structured space-time wavepackets

Ruediger Grunwald and Martin Bock

*Max Born Institute for Nonlinear Optics and Short-Pulse Spectroscopy, Berlin, Germany*

Correspondence:
Ruediger Grunwald
Max Born Institute for Nonlinear Optics and Short-Pulse Spectroscopy
Max Born Strasse 2a
12489 Berlin
Germany
*grunwald@mbi-berlin.de*

# Needle beams and structured space-time wavepackets

Abstract:

Recent research on needle beams and space-time wavepackets (STWPs) is presented. Quasi-nondiffracting STWPs propagate at stable spatial and temporal localization over extended distances. In a simple model, STWPs are interpreted as being composed of differential needle beams. Simulations indicate that pulsed Bessel-like needle beams can reach higher spectral and temporal homogeneity compared to spatio-spectrally shaped focused Gaussian beams. The interference of femtosecond needle beam arrays leads to nondiffracting self-imaging in space and time. High-speed switching of STWPs with combined optical systems, orbital angular momentum generation with self-torque and other emerging fields of research are addressed.

## 1. Needle beams

Strong localization of light in space and time is of exceptional interest for an improved understanding of light-matter interactions as well as for applications in modern quantum technologies [1]. Recently, we gave an overview on different classes of "needle beams" with particular emphasis on macroscopic quasi-nondiffracting (QND) beam structures [2]. Such beams can be applied to generate optical STWPs [3,4] which currently are the subject of intense research [5-7]. Here present a short update concerning this topic. QND-type needle beams are closely related to propagation-invariant solutions of the wave equation like Bessel beams [8] which are suitably modified by special boundary conditions. Bessel-like beams are obtained by phase shaping with axicons or programmable phase maps. To filter out the central lobes of Bessel beams at minimal diffraction, self-apodized truncation has proved to be an optimum approach [9,10]. At small angles, which can be realized by thin-film axicons or spatial light modulators, the depth of focus can exceed several 100 times the spot diameter. In contrast to Gaussian foci, a more distinct needle-like geometry is approached [11] while other advantageous properties of QND beams like self-reconstruction [1] and spectral-temporal robustness [12] remain preserved. By truncating Bessel beams at higher order minima, a further reduction of divergence is feasible [13].

## 2. Space-time wavepackets and wavepacket arrays

It was demonstrated that the needle beam approach [3,4] can be extended to more complex structured beams including highly localized wavepackets [14,15] and needle beam arrays [1,16]. In temporal and spectral domain, quasi-nondiffracting conditions are of fundamental importance by enabling for a stable pulse duration and conserved spectral bandwidth all over the nondiffracting zone. At ultrashort pulse durations, the classical picture of beams has to be replaced by propagation traces of traveling wavepackets. For tailoring structured STWPs, one has to consider space-time coupling including the complete angular-dependent pulse travel time, dispersion and diffraction effects [17-24]. The theoretical treatment of structured STWPs is not trivial because of specific constraints concerning the simultaneous control of temporal and 3D spatial parameters. Closed-form expressions with spherical wavefronts were published for rotationally-symmetric needle pulses [25]. Spatially dependent frequency control is used to compensate propagation dependent frequency shifts. The complex dynamics of ultrashort-pulsed light fields requires searching for simple and experimentally accessible approaches for inverse design strategies [26]. The representation of structured STWPs as composed of "differential" needle beams was shown to be promising [14,15].

In Figure 1, two selected design approaches of needle-shaped STWPs are compared with each other in wave propagation simulations with Python. Figure 1a shows the spatial propagation for Gaussian illumination of an axicon. Needle-shaped STWPs are concatenated in axial direction to generate an extended QND zone. The propagation zone for comparable parameters obtained by spectral management of a Gaussian focus according to ref. [27] is plotted in Figure 1b. The advantage of this approach consists in enabling, in some limits with respect to the bandwidth, to compensate for pulse broadening by dispersion.

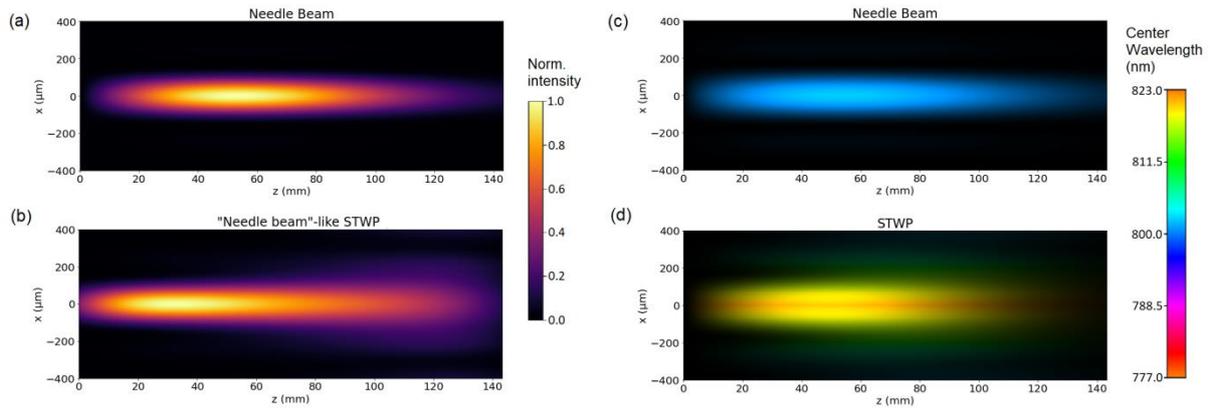

Figure 1. Needle-shaped STWPs simulated for selected design approaches: (a) spatial intensity propagation based on experimental parameters (axicon illuminated by a Gaussian-shaped laser (central wavelength 800 nm, pulse duration 20 fs, axicon angle 0.1°, refractive index 1.45, initial beam diameter 500 μm, axicon diameter adapted to beam geometry); (b) propagation of a Gaussian focus with spatio-spectral shaping according to ref. [27] for related parameters, (c),(d) spatial-spectral characteristics for approaches (a),(b), respectively (color code: relative intensity).

The propagation zones in Figures 1c and 1d indicate a higher radial homogeneity of the spectral content for the Bessel-Gaussian wavepacket compared to the spectrally confined Gaussian focus. For the chosen parameters, a transversal frequency shift of about 10 nm was found.

Further simulations were performed for structured STWPs. The interference of periodic arrays of pulsed needle beams gives rise to the nondiffracting Talbot effect where the spatio-temporal parameters experience a periodic revival both in space and time [15,28,29]. The characteristic self-imaging of a hexagonal array of a fs-scale STWP is demonstrated in Figure 2. Intensity patterns of the focal array and three subsequent nondiffracting Talbot distances are visible in radial and axial cuts in Figures 2a-c. Corresponding 2D Talbot carpets for focal distance, first out-of-phase and in-phase distances are plotted in Figures 2d-f.

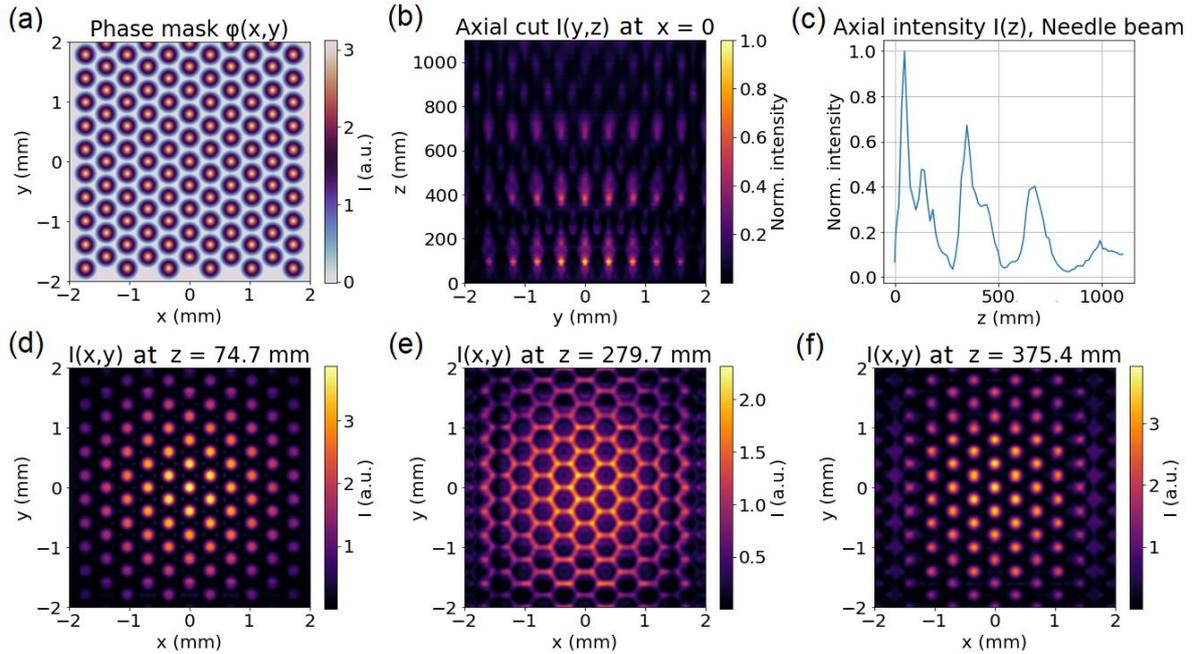

Figure 2. Self-imaging of an STWP array visualized by the simulated intensity propagation behind an axicon array (central wavelength 800 nm, pulse duration 20 fs, axicon angle 0.24°, 10 x 10 elements, period 400 μm, axicon refractive index 1.45, illuminating Gaussian beam diameter 5 mm); (a) phase mask, (b) 2D cut in propagation direction, (c) 1D central cut in propagation direction crossing the focal plane and 3 subsequent nondiffracting Talbot peaks, (d) light distribution at the focal plane, (e) out-of-phase plane. (f) 1st nondiffracting Talbot plane.

The case of aperiodic arrays (Montgomery effect) was investigated by another group [30]. Currently, emerging applications of single and multiple STWPs, e.g. arrays of pulsed orbital angular momentum (OAM) beams [31], for high-speed optical communication, real-time wavefront sensing and beam tracking [32] image transfer, or parallel materials processing [33] are studied. By combining fast switching components and reflective MEMS, high-speed adaptive shaping of structured wavepackets was obtained [34]. Moreover, we demonstrated the generation of pulsed self-torque OAM beams of propagation dependent angular velocity via radially chirped spiral grating structures programmed in a spatial light modulator [35]. Alternative methods of torquing of light and related fundamental aspects were recently presented [36].

## 3. Conclusions and outlook

To conclude, the topic of needle beams is extended by temporal and spectral degrees of freedom of ultrashort pulses and adaptive techniques. Structured STWPs can be tailored in all dimensions on the basis of nondiffracting beam shaping. By wave propagation simulations, a comparison between axicon-shaped Bessel-Gauss and a spectrally optimized Gaussian focal zone was performed. The resulting spatial structures are similar, whereas a higher radial spectral homogeneity appears for the axicon approach. The temporal homogeneity will be a subject of continuing investigations. Simulations of the nondiffracting Talbot effect confirm our former experimental results [16]. The combination of axicon shapers and spectral management via gratings could unify the specific advantages of the different approaches. The complementary relationship between QND STWPs and simultaneous spatio-temporal focusing [37] may stimulate further research on pulse front management, X-pulse propagation and rogue waves as manifestations of spectral interference in space and time. Design approaches based on the space-time duality promise to be an appropriate frame for a generalized description [38].

High-power applications like intra-cavity nondiffracting mode shaping or filament generation require to involve nonlinear dispersion and plasmonic effects. The specific behavior of nanometer and attosecond scale STWPs [39], e.g. ultrafast pulsed photonic nanojets, was not addressed here but unquestionably deserves a separate discussion. The same holds true for STWPs in fibers [40], accelerating wavepackets [41], applications of metamaterials [42], structured vector wavepackets [43], optical spatiotemporal skyrmions [44] or frequency combs [45], to mention only selected emerging topics that will further stimulate the generation, transformation and detection of spatially and temporally structured wavepackets.

## Acknowledgements

We thank Prof. N. Picqué for providing laboratory and software resources. Studies on shaping of needle beams and STWP were financially supported in parts by Deutsche Forschungsgemeinschaft (GR1782/16-1) and Laserlab-Europe (MBI 002528).